\documentclass{llncs}
\usepackage{url}
\usepackage{graphicx}
\usepackage{multirow}
\usepackage{fancyvrb}
\begin{document}
\pagestyle{empty}
\mainmatter
\title{Real-Time Notification for\\Resource Synchronization}
\author{Martin Klein\inst{1} \and Robert Sanderson\inst{1} \and Herbert Van de Sompel\inst{1} \and Michael L. Nelson\inst{2}}
\institute{
Los Alamos National Laboratory, Los Alamos USA\\
\texttt{\{mklein, rsanderson, herbertv\}@lanl.gov}
\and
Old Dominion University, Norfolk USA\\
\texttt{mln@cs.odu.edu}
}
\maketitle
%
%
%
\begin{abstract}
Web applications frequently leverage resources made available by remote web servers. 
As resources are created, updated, deleted, or moved, these applications face 
challenges to remain in lockstep with the server's change dynamics. Several approaches 
exist to help meet this challenge for use cases where ``good enough'' synchronization is 
acceptable. But when strict resource coverage or low synchronization latency is required, 
commonly accepted Web-based solutions remain elusive. This paper 
details characteristics of an approach that aims at decreasing synchronization latency while 
maintaining desired levels of accuracy. The approach builds on pushing change notifications
and pulling changed resources and it is explored with an experiment based on a DBpedia Live 
instance.
\end{abstract}
\section{Introduction}
The Web is highly dynamic \cite{cho:estimating,fetterly:evolution,ntoulas:new_on_the_web}, 
with resources continuously being created, updated, deleted, and moved. 
Digital libraries that leverage third party resources face the challenge of keeping in step 
with this rate of change. 
There are significant use cases in digital libraries that require real-time and accurate 
synchronization, and in many cases this need is addressed through ad-hoc technical approaches
implemented by a small group of collaborating systems. Proposals for more generic, Web-scale, 
synchronization approaches have been suggested but have not been widely adopted. 

ResourceSync, a project launched by the National Information Standardization 
Organization\footnote{\url{http://www.niso.org}} (NISO) and the Open Archives 
Initiative\footnote{\url{http://www.openarchives.org}} (OAI) aims to design an 
approach for resource synchronization that is aligned with the Web 
Architecture \cite{web_architecture} and that has a fair chance of adoption by 
different communities. The ResourceSync effort recognizes the challenge of devising an 
approach that can be applied across a variety of use cases that entail different 
types of resources, different types of changes, and differing requirements regarding the 
coverage and speed of synchronization. 
However, the current specification \cite{klein:resync_spec}
is focused on a pull-based synchronization approach and does not consider push-based
implementations. 

In this paper we present a design for a push-based approach as an extension of the Resourcesync
framework and investigate its performance for a synchronization 
use case that entails resources that change at a very high frequency. 
We synchronize two systems with an instance of DBpedia Live \cite{dbpedia_live}
via our prototype push-based architecture and measure synchronization latency and
accuracy. 
This use case is representative for modern digital libraries that rely on Linked Data 
to integrate constantly changing resources from various datasets. For example,
the BBC Linked Data applications integrate data from Last.FM, DBpedia, MusicBrainz, GeoNames,
and others.
\section{Related Work} \label{sec:Related Work}
Resource synchronization is not a new problem. It has been acknowledged as an issue, among others, by 
Tim Berners-Lee \cite{tbl:delta}, Umbrich et 
al. \cite{umbrich:dataset_dynamics}, and the related W3C Dataset Dynamics 
activity\footnote{\url{http://www.w3.org/wiki/DatasetDynamics}}.
The herein presented work is embedded in the ResourceSync project, which has published a specification
for Web resource synchronization \cite{klein:resync_spec} (a beta draft at the time of writing).
The theoretical foundations of the ResourceSync framework were detailed in \cite{vdsomp:resync_dlib}
and its technical aspects were described in \cite{klein:resync_dlib}.
The below overview of work in the problem domain distinguishes between approaches that rely on optimizing 
pull interactions with resources, those that are based on communicating change notifications, and others 
that focus on the transfer of changed content only.
\subsection{Push/Pull Approaches}
Synchronization approaches are based on a pull method, a push method, or a hybrid of both. 
The issue of which approach to use under which circumstances has been the subject of several research endeavors. 
For example, Bhide et al. \cite{bhide:adaptive_push-pull} theoretically compare push and pull approaches for 
disseminating dynamic web data with an emphasis on a client's temporal coherency requirement. They introduce 
\textit{Push and Pull (PaP)}, a method with a scheduled pull being the default mechanism but where the server 
also has the capability to push changes if it foresees that the polling client would miss the changes otherwise.
Another concept they discuss is \textit{Push or Pull (PuP)} where the push method is the default and the server 
can dynamically allocate push or a pull channels to clients, depending on available resources.
Silberstein et al. \cite{silberstein:feeding_frency} explored the boundary conditions where it was optimal to 
push or pull feeds, depending on frequency of reads and writes. Their work suggests that in order to setup a 
resource-aware synchronization implementation local decisions on a creator/consumer basis are necessary. In 
this sense, they show that the push method is preferable if the user's consumption frequency is greater than the 
event creation frequency and pull otherwise.
%
%
%
%
\subsection{Change Notifications}
The Event Notification Protocol \cite{enp} specifies requirements for change notifications in relation to WebDAV implementations. 
\textit{DSNotify} introduced by Popitsch and Haslhofer \cite{popitsch:dsnotify} is a change detection and notification 
framework for Linked Datasets. 
Volz et al. \cite{volz:links_web_of_data} introduced \textit{Silk}, a linking framework based on the Web of Data Link 
Maintenance Protocol\footnote{\url{http://www4.wiwiss.fu-berlin.de/bizer/silk/wodlmp/}}.
Ping the Semantic Web\footnote{\url{http://pingthesemanticweb.com/}} offers change notification as a web service.
All three systems are geared towards Linked Datasets and are not designed as generic resource synchronization frameworks. 
They further rely on aggregated baseline data to be available in a central service, raising scalability concerns in light 
of an ever expanding Linked Data cloud. 
\emph{Pingback} and its extension \emph{Semantic Pingback} \cite{tramp:semantic_pingback} both provide a lightweight 
notification approach. However, the approaches are based on subscriptions to individual resources, making them problematic 
for large resource collections. 
The HTTP-based Simple Update Protocol (SUP)\footnote{\url{http://code.google.com/p/simpleupdateprotocol/}} and the UDP-based 
Simple Lightweight Announcement Protocol (SLAP)\footnote{\url{http://joecascio.net/joecblog/2009/05/18/announcing-slap/}}
provide conceptual mechanisms to notify about change events. However, they both suffer from the lack of acceptance and reference 
implementations.
\section{Architectural Paradigms} \label{sec:Architectural Paradigms}
At the core of the resource synchronization problem is the need for one or more Destination servers to
remain synchronized with (some of the) resources made available by a Source server.
For typical synchronization scenarios we can distinguish between a change notification (CN) requirement, 
which allows a Destination to understand that a resource has changed at the Source and what the type of 
change is (e.g., update, delete) and a content transfer (CT) requirement, which allows a Destination 
to update its holdings to reflect the change the resource underwent at the Source.

Pull-based approaches are inarguably the simplest alternative 
for resource synchronization as Destinations recurrently poll the Source for changes and retrieve 
resources that have to be synchronized. However, these approaches are not subject of this paper.
They have been explored in our previous work \cite{vdsomp:resync_dlib} and implemented in the 
ResourceSync framework \cite{klein:resync_spec}.
Therefore, in this section we consider two push-based synchronization architectures that adhere to the 
separation between CN and CT.
\subsection{CN Simulated Push \& CT Pull}
\begin{figure}[t]
\center
\includegraphics[scale=0.25]{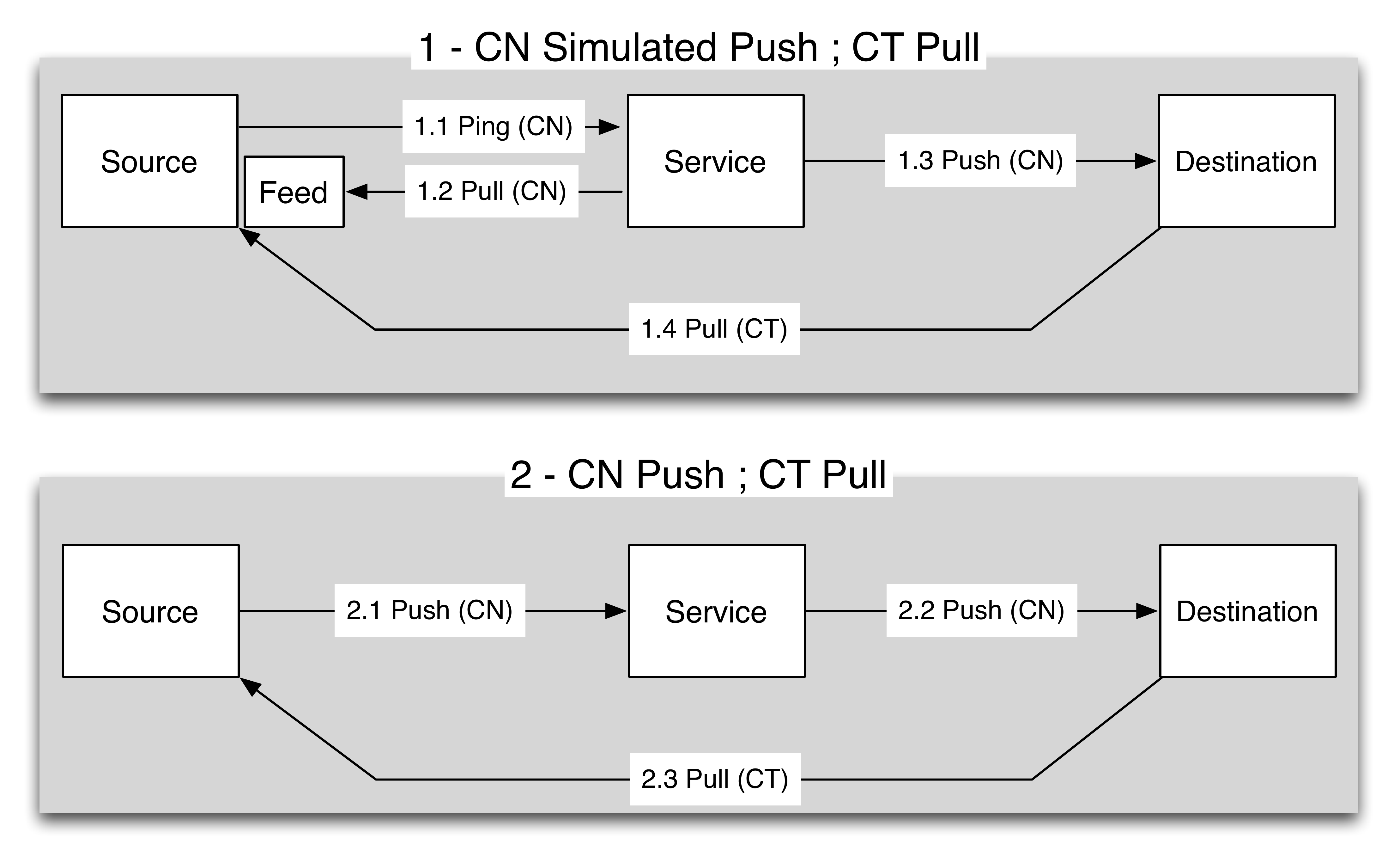}
\caption{Architectures Paradigms}
\label{fig:arch}
\end{figure}

Introducing feed-based technology (e.g., Atom/RSS, OAI-PMH repository) at the Source was a first step to 
improve pull-based approaches. This technology introduces a new component into the synchronization 
architecture other than Source and Destination: a Service. More recent approaches, such as PubSubHubbub 
(PuSH)
, have decoupled such a Service from the 
Source, reducing the memory burden the latter carries but 
adding interactions. Such an architecture is displayed in $1$ of Figure \ref{fig:arch}.

Typical PuSH use cases are concerned with the transfer of Atom records, commonly conveying news items, 
from publishers to subscribers. But PuSH could also be applied to transfer CNs in 
which case a feed is maintained by the Source. When the Source's feed content is updated (i.e., one 
or more resources changed), the Source sends a content-less ping to a Hub Service 
(CN interaction $1.1$). The Hub Service then retrieves the Atom feed from a previously registered location 
(CN interaction $1.2$) and consecutively uses AtomPub to push Atom entries to Destinations (CN 
interaction $1.3$). In order to make this possible, Destinations have previously subscribed to the Hub 
Service and have provided a callback URI. Upon receipt, the Destination pulls changed resources, 
if required (CT interaction $1.4$).
Since interactions $1.1$ and $1.2$ are a ping and a pull (but not a push), we refer to this approach as 
simulated push.

From the perspective of the Source, this architecture is convenient as it 
moves the burden of maintaining a longer-term change notification memory to the Hub Service. 
The Hub maintains a feed per Source and, per feed, keeps track of each Destination's callback URI.
%
The initial ping is easy to implement for the Source, and reduces unnecessary interactions. However, without 
the introduction of special-purpose optimizations 
the Hub will repeatedly retrieve the same CN entries. The requirement in PuSH for the Destination to have a 
web presence for callback may not be appropriate or even possible in certain use cases, for example when the 
Destination resides on a mobile device.

A scenario we call selective synchronization, refers to a Destination's need to only remain in sync 
with a subset of resources made available by a Source, or to only be aware of certain types of its changes.
The notion of a channel as a conduit for notifications about a subset of a Source's changes
follows quite naturally. An example could be a Source defining channels to reflect DBpedia categories.
However, for selective synchronization, simulated push is problematic. 
The Source would have expose multiple feeds and the Hub Service must maintain 
each of them. Also, it is likely that Destinations have to maintain one callback per channel they are 
interested in. If the Source has many such channels, and if resources are represented in multiple 
channels, the number of interactions involved (CN interactions $1.1$ and $1.2$) increases significantly.
\subsection{CN Push \& CT Pull}
A real, rather than a simulated, push approach for CNs ($2$ in Figure \ref{fig:arch})
is the focus of this paper.
It consists of the Source pushing CNs via the intermediation of a Service to Destinations (CN interactions 
$2.1$ and $2.2$). As in the previous approaches, upon receipt of a notification, the Destination may decide 
to obtain the changed resource (CT interaction $2.3$).

Interestingly, the memory requirements for the Source are zero: it can fire off CNs and 
immediately forget about them. The number of transactions is reduced as the initial ping of the previous 
architecture is no longer required, but the real optimization comes in terms of the Service not having to 
pull the Source's feed(s) for every change. The payload for the CNs could be designed to meet synchronization
requirements, rather than conveying them in a less than suitable container like Atom entries.

An example technology in this space is XMPP \cite{rfc:6120,saint-andre:xmpp}, used for instant messaging 
between humans but also for machine to machine communication \cite{xep:lop}. This technology 
requires the Destination to listen when notifications are transmitted; this is not dissimilar to being online
for the callback method of PuSH. XMPP has its own protocol control structure in XML and is approximately the
same order of magnitude in bytes as the Atom structure, however redundant elements are not passed backwards 
and forwards reducing required bandwidth.

The XMPP PubSub extension \cite{xep:60} provides supports for channels, with a Service rather than both 
Service and Source in charge of maintaining them. The Source publishes the messages to appropriate channels, 
which it can create as desired. The Service passes these messages on to Destinations that subscribe to the 
channel. XMPP also has an extension that allows the nesting of channels so that a subscriber to a channel 
receives messages from all of its child channels. For a large and dynamic set of resources, this capability 
significantly reduces unnecessary transactions.
\section{Experiment} \label{sec:Experiment}
\begin{figure}[t]
\centering
\includegraphics[scale=0.2]{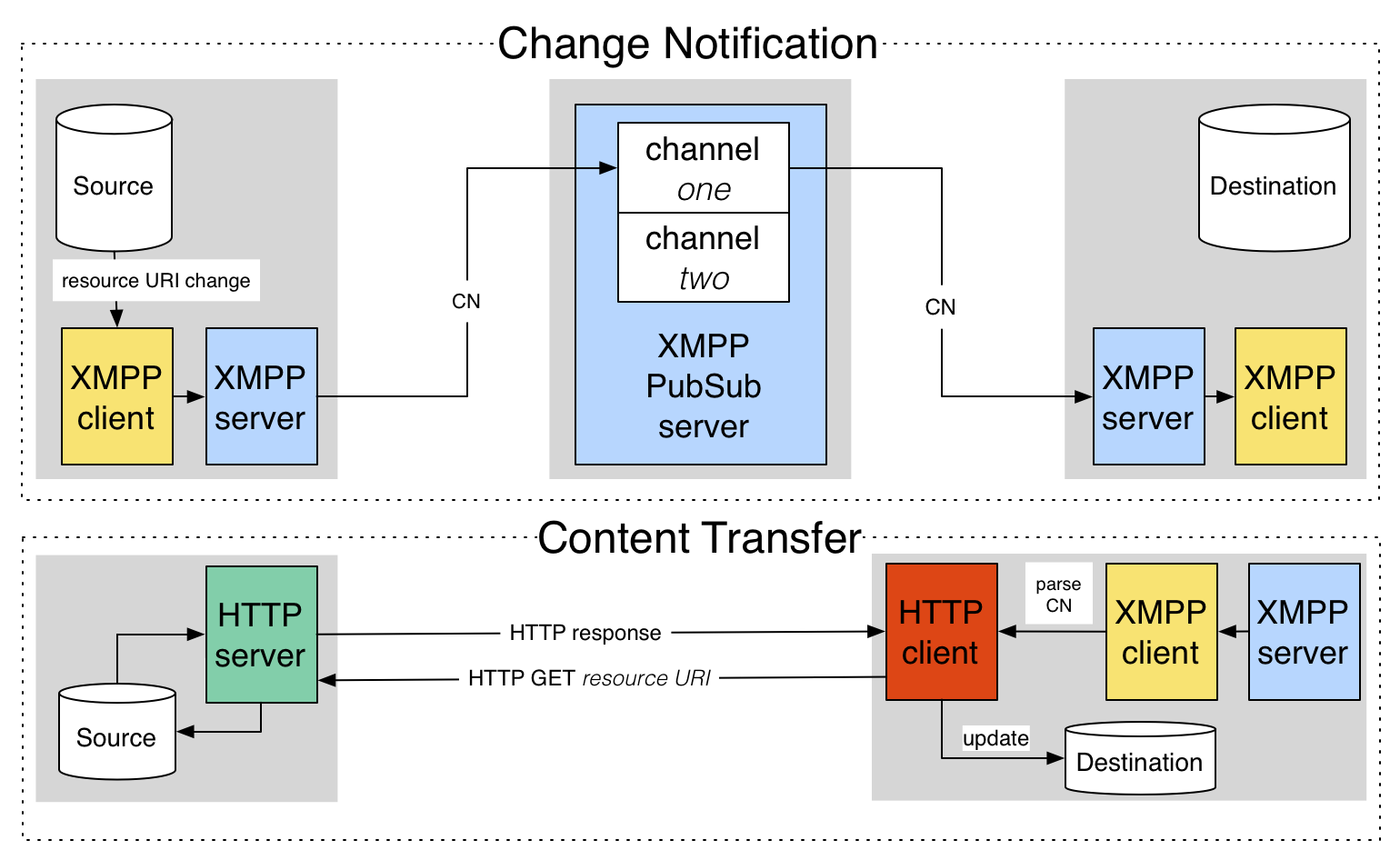}
\caption{An architecture based on XMPP PubSub}
\label{fig:our_architecture}
\end{figure}
In order to explore the problem domain, an example resource synchronization approach was formulated and tested. 
The approach consists of pushing CNs from a Source to Destinations. 
The format and content of the CNs are not subject of this experiment. For a detailed discussion, please refer to 
our previous work \cite{klein:resync_dlib}.
CT consists of a Destination pulling the entire resource about which a Destination received a change notification. 
The approach was tested to synchronize two Destinations with a Source that is a DBpedia Live instance operated at 
the Los Alamos National Laboratory (LANL).

There are several motivations for these choices. 
First, little is known regarding reliability and scalability of 
change notification approaches, as recognized in \cite{umbrich:dataset_dynamics}.  Also, the considerable update 
frequency of DBPedia Live topic-URI graphs, which was observed to average around two changed graphs per second 
during a two month period, provides a credible synchronization challenge. Finally, intuition suggests that such 
high change frequency may cause problems, especially for an approach where entire resources rather than only 
changes are communicated. 
\subsection{Synchronization Approach}
The tested approach (Figure \ref{fig:our_architecture}) handles CNs by means of a 
push paradigm. XMPP PubSub is used as the protocol to push notifications from Source to Destinations. 
Consequently, Source and Destination each have a Jabber ID, operate an XMPP client to send/receive notifications, 
and have an XMPP home server. XMPP PubSub nodes serve as channels for selective synchronization. 
The Source creates the channels and publishes notifications to them. 
Destinations subscribe to channels and as a result receive notifications.
The XMPP PubSub server manages channels and subscriptions, and relays CNs from Source to Destinations. 
CT is handled by means of a pull of the entire changed resource. When a Destination 
receives a CN, it determines whether it should obtain the corresponding updated version of the resource. If 
so, it dereferences that resource's URI. Note that, when a resource changes very rapidly or when the Destination 
processes CNs with a delay, the possibility arises that the resource version pulled by the Destination is more 
recent than the event time reported in the CN.
\subsection{Configuration}
\begin{figure}[t]
\centering
\includegraphics[scale=0.24]{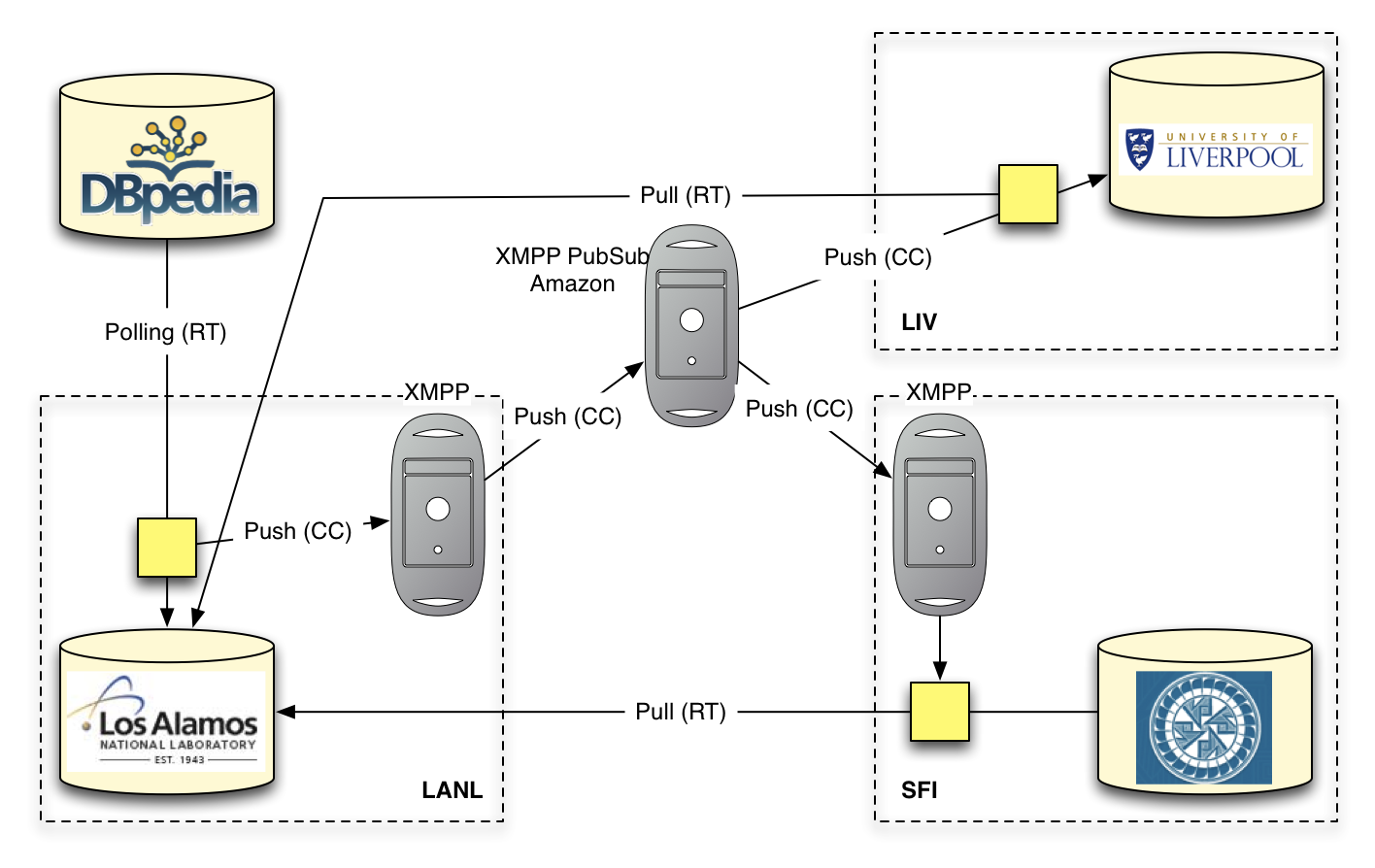}
\caption{Systems involved in the experiment}
\label{fig:experiment_setup}
\end{figure}
\begin{figure}[t]
\centering
\includegraphics[scale=0.3]{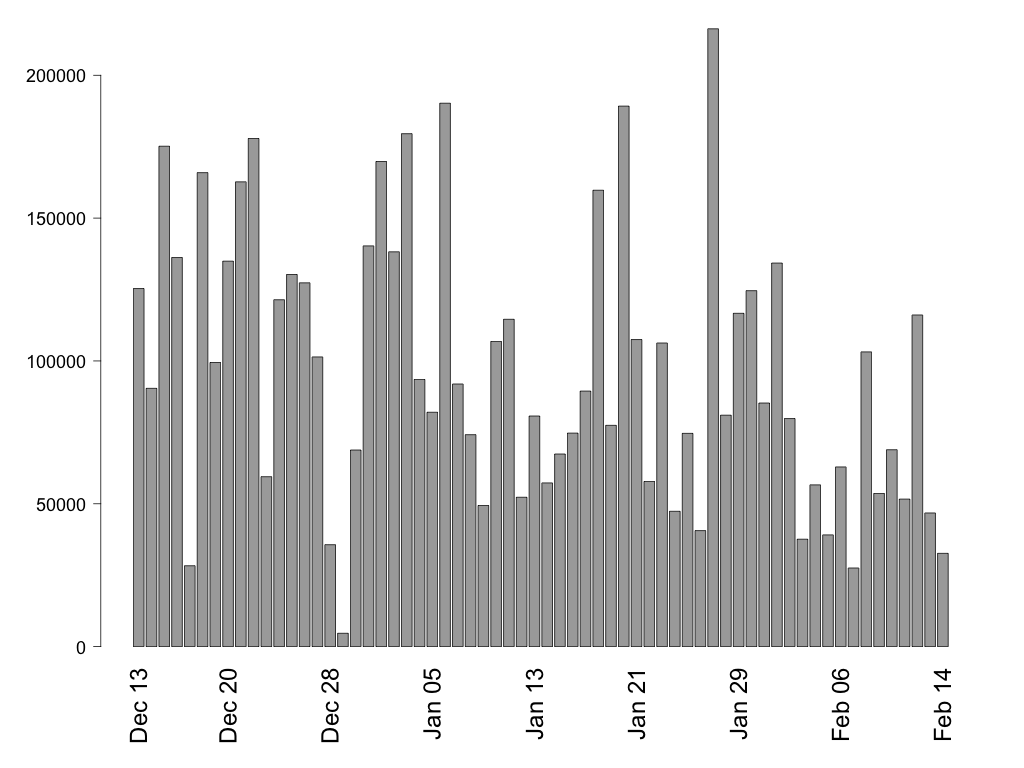}
\caption{CNs on dbpedia\_all per day}
\label{fig:updatefrequency}
\end{figure}

To create a baseline for synchronization, the LANL DBpedia Live instance (bottom left of Figure \ref{fig:experiment_setup}) 
uploaded a dump from DBpedia Live (top left of Figure \ref{fig:experiment_setup}). 
Since then, it has kept in step with changes using the latter's content-type-specific changeset 
mechanism\footnote{See \url{http://live.dbpedia.org/liveupdates/}} that consists of providing an updated and deleted 
changeset file per update cycle, in which each line is a triple that has a DBpedia Live topic-URI as subject.  
The LANL instance continuously monitors the availability of new changesets and, when available, processes them. The 
heuristics involved in doing so are beyond the scope of this paper. It suffices to say that at the end of each processing 
cycle, the LANL instance is updated through the addition of new versions of previously existing topic-URI graphs, the 
deletion of topic-URI graphs, or the creation of new ones. Each of these events is communicated on the \texttt{dbpedia\_all} 
channel that covers all changes to all DBpedia topic-URIs. Figure \ref{fig:updatefrequency} shows the resulting daily 
traffic. Overall, 99\% of these CNs are about updates, 0.6\% about deletions, and 0.03\% about creations. In 
addition, depending on the DBpedia category of a topic-URI, CNs are also sent to channels such as \texttt{dbpedia\_music}, 
\texttt{dbpedia\_business}, etc. The LANL instance holds on to all versions of topic-URI graphs and stores its 
serializations.
%
%

An XMPP PubSub server operated in Amazon's cloud (center of Figure \ref{fig:experiment_setup}) provides the infrastructure 
for the channels. The LANL instance creates the channels to which Destinations can subscribe. In 
the experiment, two 
remote servers subscribe. A server at the University of Liverpool (LIV, top right of 
Figure \ref{fig:experiment_setup}) subscribes to dbpedia\_all. Upon receipt of the CNs, it updates its local DBpedia 
collection by dereferencing or deleting the corresponding topic-URIs. It stores serialized graphs in its file system in the 
same way the LANL instance does. A server at the Santa Fe Institute (SFI, bottom right of Figure \ref{fig:experiment_setup}) 
remains synchronized in the same way, but its topic-URI collection is maintained in a Virtuoso triple store. Neither of 
these Destinations performed a baseline synchronization and they do not hold on to old versions. Each runs an XMPP client 
that connects to a local XMPP server. 

Figure \ref{fig:experiment_setup} depicts the information flow in the experimental setup: the LANL instance is polling 
DBpedia Live for CT; the LANL instance sends CNs about changes it is undergoing and these CNs are relayed by 
the XMPP PubSub server in the cloud to the Liverpool and Santa Fe servers. These servers then use HTTP GET to obtain 
new or updated resources (CT).
\subsection{Results}
\begin{figure}[t]
\centering
\includegraphics[scale=0.25]{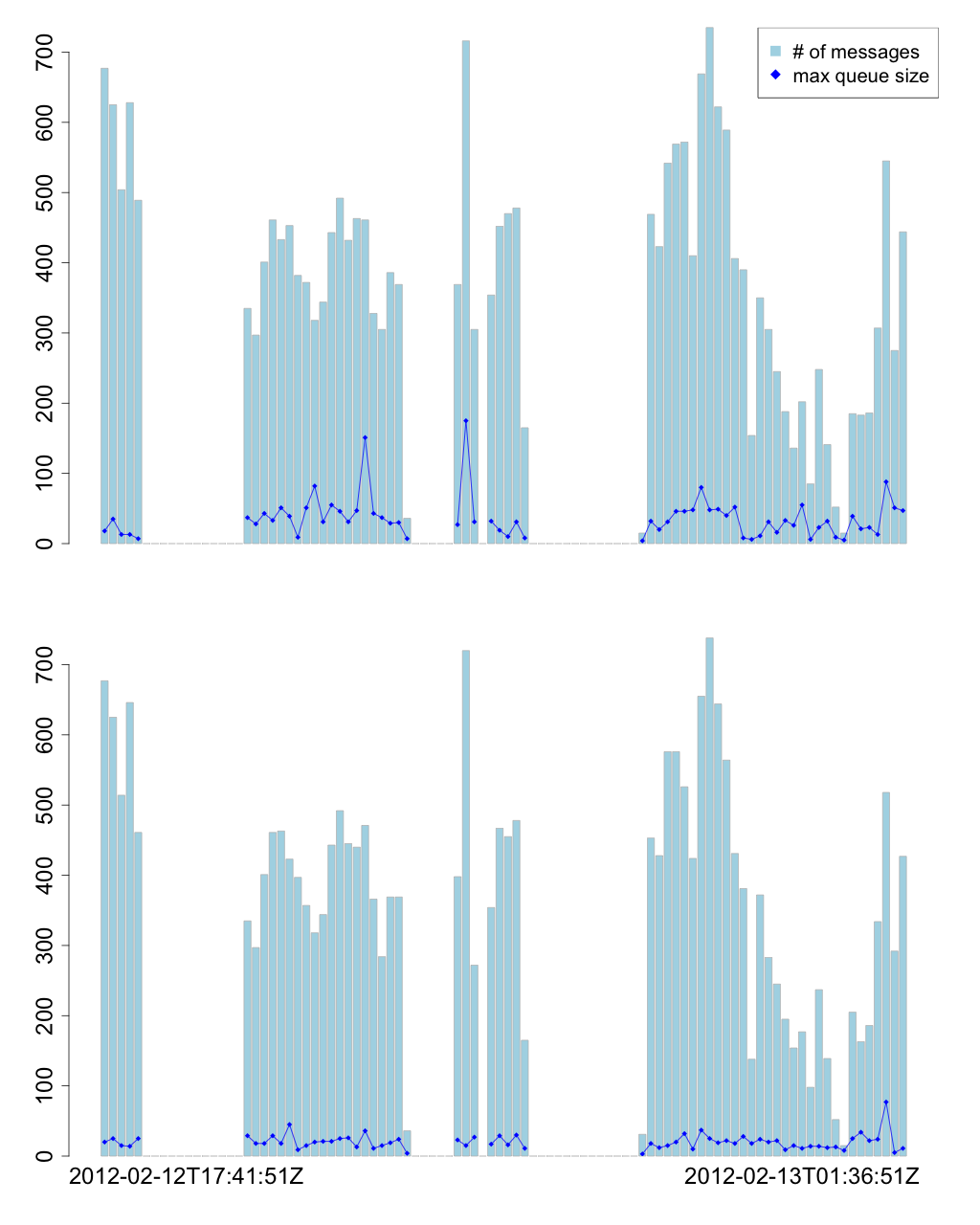}
\caption{Run 4, LIV (top), SFI (bottom) Destinations}
\label{fig:uk4}
\end{figure}

\begin{table}[t]
\centering
\begin{tabular}{|r|c|c|c|c|} \hline
\multirow{2}{*}{\textbf{Run}} & \multirow{2}{*}{\textbf{Total CNs}} & \textbf{Diff} & \multicolumn{2}{c|}{\textbf{MaxQ}} \\
&&\textbf{LANL-LIV} & \textbf{LIV} & \textbf{SFI} \\
\hline
1 & 13,819 & 4 (0.03\%) & 120 & 57 \\
2 & 32,453 & 2 (0.006\%) & 389 & 65 \\
3 & 6,910  & 1 (0.01\%) & 93 & 44 \\
4 & 24,400 & 5 (0.02\%) & 175 & 77 \\
5 & 11,850 & 0 (0.0\%) & 84 & 57 \\
6 & 14,937 & 4 (0.03\%) & 98 & 61 \\
\hline
\textbf{total} & 104,369 & 16 (0.015\%) & & \\
\hline
\end{tabular}
\caption{Experimental Results}
\label{tab:results1}
\end{table}
A first experiment investigated whether the Destinations at LIV and SFI were able to keep up with the pace of CNs
sent out by LANL. As CNs arrived at the Destinations, they were pushed into a first-in-first-out queue awaiting 
processing. This entailed obtaining the changed resource from LANL in case of create/update events and updating the 
local versions for all events. The size of this queue as compared to the amount of received CNs serves as an 
indication of the capability of a Destination to remain in lockstep. These parameters were observed during six 
eight-hour runs conducted at different times of the day, in the course of a single week. During these sessions, 
the output rate of changesets from DBpedia Live was very inconsistent. As a result, LANL sometimes broadcast 
thousands of CNs in a single five minute interval, and at other times not a single one for several hours. Even in 
less extreme situations, CNs were sent out in a bursty manner as a result of LANL polling DBpedia Live
every $30$ seconds for changesets but pushing out all resulting notifications once those were processed. Due to this 
bursty sending, the queues at the Destinations could grow rapidly. Nevertheless, both Destinations were able to keep 
up with the barrage of CNs. The second column in Table \ref{tab:results1} shows the total number of CNs processed 
per run, whereas the ``MaxQ'' columns show the maximum queue sizes for each run for LIV and SFI, respectively. As can 
be observed, LIV had significantly larger queues than SFI; this is due to additional time required to dereference resources 
from a location on another continent. Figure \ref{fig:uk4} depicts the number of CNs over the $8$ hour session of 
the fourth run measured at five minute intervals, as well as the maximum queue size during each interval. It needs to 
be noted that for all runs, both LIV and SFI ended up with an empty queue at the end of each of these five minute intervals. 

A second experiment investigated whether resources were effectively synchronized. This experiment was conducted during the 
same six eight-hour runs. It involved LANL as the Source and LIV as the Destination because both use the same storage 
mechanism for topic-URI graphs, allowing for a byte-level comparison of the respective collections. Lack of control over 
the incoming data from DBpedia Live caused a problem with dereferencing and hence synchronizing some of the 
communicated URIs. For example, some DBpedia Live URIs contained the same character both percent-encoded and 
unencoded, and the LANL storage layer was unable to resolve such identifiers successfully. Occurrences of this problem 
are excluded from the results, but amounted to significantly less than $0.1\%$ of all CNs. Note that this problem 
results from LANL having to mint and communicate URIs that contain parts of the DBpedia Live in their path, 
and would 
not occur in a real-life situation. A recursive \texttt{diff} of the Source and Destination collections was used at the end of 
each run to detect resources that were not synchronized correctly. 
The ``Diff LANL-LIV'' column in Table \ref{tab:results1} shows the number and percentage of resources at LIV that were 
out-of-sync with LANL at the end of each run. 
Overall, 99.99\% of the synchronization was accurate after processing more than 100,000 CNs.

To verify the intuition that the changeset approach used by DBpedia Live would naturally result in  a 
smaller content 
transfer payload, a third experiment was conducted. During a $24$ hour period, the sizes were monitored of respectively 
all changesets pulled by LANL from DBpedia Live, and all HTTP GET transactions issued from SFI to LANL that 
resulted from processing CNs. 
The changesets amounted to $149$MB compressed, which corresponds to $4.2$GB uncompressed. The HTTP GET transactions totaled 
$1.8$GB uncompressed. Assuming these transactions would have had compressed response bodies, and applying a reasonable 
compression coefficient of about $0.1$ as shown in \cite{fernandez:rdf_compression}, HTTP GET transactions would have totaled 
around $180$MB. 
\section{Conclusions} \label{sec:Conclusions}
This paper contributes to the exploration of Web-based resource synchronization approaches.
It demonstrates the real possibility of using an XMPP PubSub architecture to push CNs
and use a pull-based method for CT. 

Clearly additional experiments are required, but results from the reported work were far beyond expectations. 
A more robust implementation, without the described URI encoding issues and without the need to poll 
DBpedia Live in order to update LANL, could likely achieve $100\%$ synchronization without significant 
additional implementation effort. The extent of unnecessary messages is minimized thanks to the intermediation 
of Service-side channels. The synchronization latency, even under extreme load, remained very low and can 
further be decreased by adding additional retrieval threads. Generally, even though developers were not previously 
acquainted with XMPP protocols or tools, adoption was straightforward and off-the-shelf client and server tools were 
readily usable. 

The tested approach relies on URI dereferencing for content transfer. Even though the experiment did not 
encounter specific problems with this regard, the following issues arose when discussing the reported work. Firstly, 
some resources, like those identified by hash-bang HTTP URIs\footnote{See \url{http://www.jenitennison.com/blog/node/154}}, 
require client-side processing to obtain a meaningful representation. Such resources introduce additional 
synchronization challenges and suggest that resources to be initially considered for a synchronization framework 
need to be cacheable\footnote{See \url{http://isolani.co.uk/blog/javascript/BreakingTheWebWithHashBangs}}. 
Secondly, specific representations of a resource are subject to synchronization. But the possibilities with 
this regard seem limited to representations that can be requested using the URI's protocol parameters. This 
introduces challenges related to the subjectivity of observed changes in case of personalized (e.g., geo-dependent) 
representations. 
\section{Acknowledgments}
We thank Simeon Warner (Cornell University), Bernhard Haslhofer (University of Vienna) and Carl Lagoze 
(Michigan University) for project input and Lyudmila L. Balakireva and Harihar Shankar of the LANL Prototyping 
Team for implementations. 
Thanks to NISO's Todd Carpenter, Nettie Lagace and Peter Murray for ResourceSync support.
This work is partly funded by the Sloan Foundation and the Library of Congress. Many thanks to the 
University of Liverpool and the Santa Fe Institute for providing computing resources.
%
%
\bibliographystyle{abbrv}
\bibliography{references}
%
%
%
\end{document}